\documentclass{article} %
\usepackage{arxiv_style,times}
\arxivcopy

\usepackage{array}
\usepackage{longtable}

\usepackage[table]{xcolor}\usepackage{graphicx}
\usepackage{amssymb}
\usepackage{makecell}
\usepackage{multirow}

\title{Graph-Guided Repository Environment Construction}
\author{
Jianying Pan\textsuperscript{1,*}, John Zhang\textsuperscript{2,*}, Hongyu Zhang\textsuperscript{1,\textdagger}\\
\normalfont\textsuperscript{1}Chongqing University\\
\normalfont\textsuperscript{2}University of New South Wales
}

\newif\ifshowcomments
\showcommentsfalse

\definecolor{jzcolor}{RGB}{120,50,150}

\usepackage{capt-of}
\usepackage{float}
\usepackage{placeins}
\usepackage{wrapfig}    %
\usepackage{booktabs}   %
\usepackage{tabularx}   %
\usepackage{listings}   %
\usepackage{hyperref}
\hypersetup{
  pdftitle={Graph-Guided Repository Environment Construction},
  pdfauthor={Jianying Pan, John Zhang, Hongyu Zhang}
}
\usepackage{url}
\usepackage{xurl}
\usepackage[normalem]{ulem}

\lstdefinestyle{promptstyle}{
  basicstyle=\ttfamily\small,
  breaklines=true,
  breakautoindent=true,
  breakindent=20pt,
  columns=fixed,
  keepspaces=true,
  frame=none,
  numbers=none,
  aboveskip=1em, belowskip=1em,
}

\begin{document}

\maketitle
\fancyhead{}
\begingroup
\renewcommand{\thefootnote}{\fnsymbol{footnote}}
\footnotetext[1]{Equal contribution.}
\footnotetext[2]{Corresponding author.}
\endgroup

\begin{abstract}

Coding agents now increasingly rely on execution to validate their solutions, making the construction of reliable execution environments a critical enabling capability. However, repository environment construction is challenging because execution requirements are fragmented across repository artifacts and may only become apparent during execution. Existing agent-based approaches address this problem through iterative interaction,
but information about the current construction state, including discovered
requirements, satisfied and unresolved prerequisites, and their dependencies,
can remain distributed across the interaction history.
We present \textsc{Graph2Env}, an agent-based approach centered on DepGraph, a typed dependency graph that explicitly represents the environment requirements needed for repository execution, their dependency relations, and their states. Graph2Env uses DepGraph to guide environment construction and continuously refines it with execution feedback, while persisting successful repairs into a replayable construction procedure. The resulting artifacts are then applied in a fresh environment to verify that the constructed environment can be reproduced. We evaluate Graph2Env on a benchmark of 200 Python repositories drawn from RATBench and EnvBench, against a static dependency-inference baseline (pipreqs), three specialized environment-construction systems (Repo2Run, RAT, and SetupX), and two general-purpose coding agents (SWE-agent and Claude Code). Graph2Env achieves an 81.0\% Environment Build Success Rate (EBSR) and a 59.3\% Environment Setup Success Rate (ESSR), outperforming the strongest
baseline by 9.5 and 9.0 percentage points, respectively.
\end{abstract}

\section{Introduction}
\label{sec:introduction}

Coding agents can now navigate repositories, edit source files, execute commands, run tests, diagnose failures, and iteratively refine their solutions. Benchmarks such as SWE-bench evaluate such agents on real-world GitHub issues by requiring them to generate patches that resolve repository-level problems~\citep{jimenez2024swebench}. Critically, many of these capabilities rely on execution: an agent must be able to run software and its tests to obtain feedback, validate changes, and determine whether a task has actually been solved. Thus, as coding agents become increasingly execution-grounded, constructing an executable repository environment becomes a fundamental enabling capability for autonomous software development.

Unfortunately, constructing such environments remains challenging. A repository rarely specifies its complete execution requirements in a single place. Instead, %
requirements for constructing a repository environment can span multiple layers, including language runtime, system libraries, build tools, and runtime configurations. Empirical studies~\citep{eliseeva2025envbench,hu2025repo2run} have repeatedly shown that automatically rebuilding software repositories is difficult and has substantial build-failure rates.
The problem is further complicated because some requirements are latent: they become observable only during execution. For example, installing a language-level package may expose a missing native compiler; resolving that compiler may reveal a missing system library; successful test collection may subsequently expose a required service or runtime configuration. Consequently, repository environment construction is an iterative process in which the system must repeatedly infer requirements, satisfy their prerequisites, execute the repository, interpret failures, and revise its understanding of the environment. 

\begin{figure}[t]
  \centering
  \includegraphics[width=\linewidth]{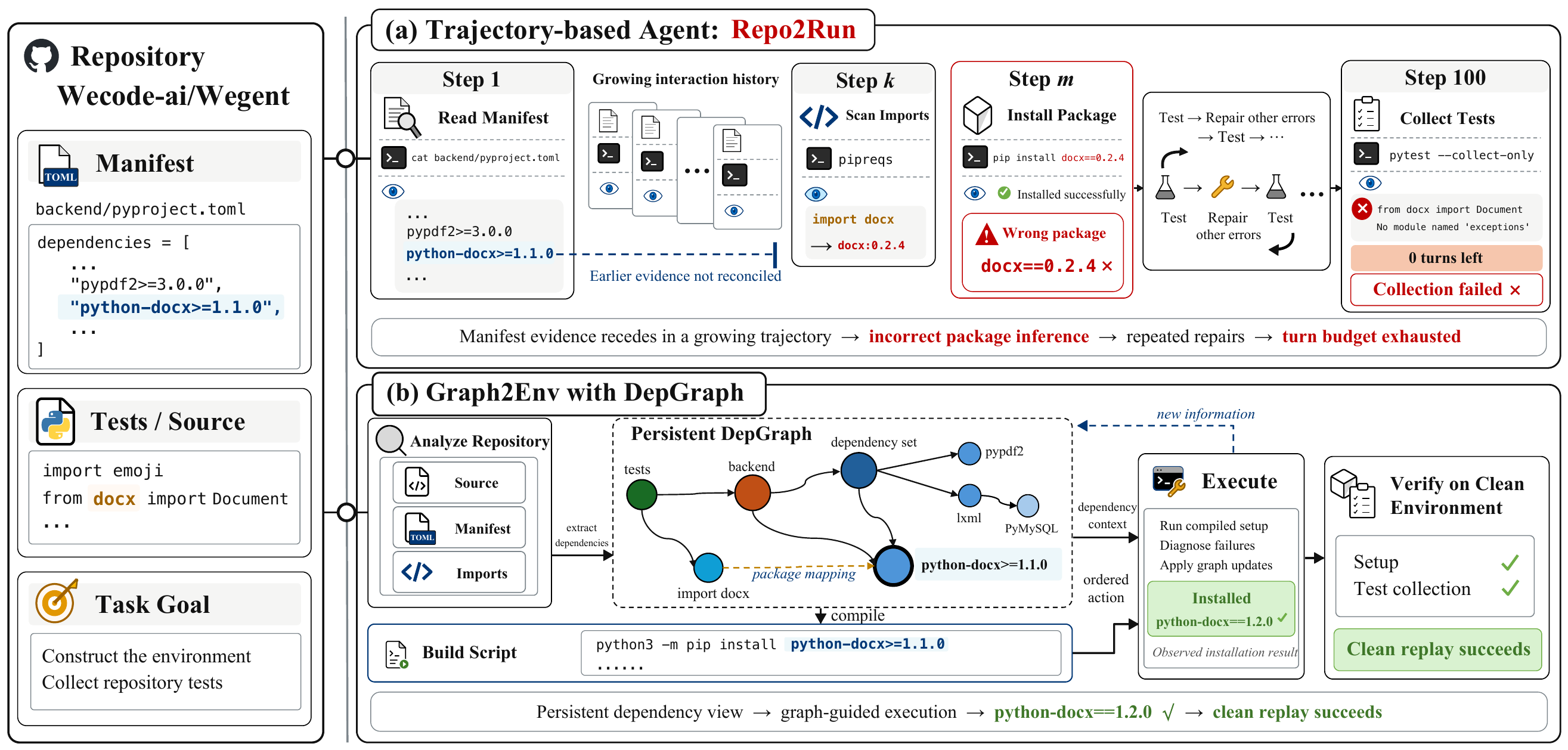}

  \stepcounter{footnote}
  \caption[Motivating comparison between trajectory-based environment construction and Graph2Env.]{
  Motivating comparison between trajectory-based environment construction and
  Graph2Env on the \texttt{wecode-ai/Wegent} repository%
  \protect\footnotemark[\value{footnote}].
  Repo2Run loses track of earlier manifest evidence and installs the wrong
  \texttt{docx} package, whereas Graph2Env preserves the import-to-package
  mapping in DepGraph, compiles it into the build script, and succeeds under
  clean replay.}
  \label{fig:motivating-example}
  \vspace{-12pt} %
\end{figure}

Recognizing this challenge, recent approaches such as Repo2Run, RAT, and SetupX~\citep{hu2025repo2run,huang2026rat,zhou2026setupx} have begun to employ LLM agents to automate repository environment construction. For example, Repo2Run uses an LLM-based build agent to automate dependency management, conflict resolution, Python-version management, and Dockerfile construction. These approaches demonstrate that agentic reasoning and execution feedback can substantially improve upon purely static dependency extraction. General-purpose coding agents such as Codex, Claude Code, and SWE-agent can also inspect repository artifacts, execute shell commands, install dependencies, observe failures, and iteratively repair their working environments. 
However, current approaches introduce a fundamental challenge: 
information about discovered requirements, satisfied prerequisites, unresolved problems, and previous repairs is largely distributed across the interaction trajectory (i.e., the accumulated sequence of actions and execution observations). As trajectories become longer, the agents must repeatedly reconstruct from their history what has already been established, what remains missing, and how individual requirements depend on one another. %
In our motivating example shown in Figure~\ref{fig:motivating-example}, this results in earlier manifest evidence being overlooked and subsequent repairs being guided by an incorrect package inference.

We argue that repository environment construction should be stateful. Rather than treating construction as trajectory-driven trial and error, an agent should maintain an explicit representation of the construction state: what requirements exist, how they depend on one another, which requirements have been satisfied, and which remain unresolved. Such a representation should not merely serve as passive memory. It should actively determine what construction actions need to be performed, evolve as new evidence is obtained from execution, and ultimately produce a procedure that can reconstruct the discovered environment from scratch.

\footnotetext[\value{footnote}]{\url{https://github.com/wecode-ai/Wegent}}

Based on this insight, we present \textbf{Graph2Env}, a stateful approach to repository environment construction built around an explicit Dependency Graph (DepGraph). DepGraph represents the execution target and heterogeneous environment requirements (imports, packages, runtimes, tools, system libraries, etc.) as typed nodes, while edges capture dependency and mapping relationships among them. %
Unlike a conventional package-dependency graph, DepGraph is an execution-grounded, evolving representation of environment-construction state.
Graph2Env first constructs an initial DepGraph from repository evidence, dependency resolution, and lightweight probing, and compiles the graph into a dependency-ordered construction procedure. During execution, an agent diagnoses failures and updates the DepGraph with newly discovered or revised requirements and dependencies, triggering recompilation when necessary. Finally, Graph2Env replays the resulting procedure from a clean base to verify reproducible environment construction.

To evaluate Graph2Env, we construct a new evaluation benchmark of 200 Python
repositories drawn from RATBench and EnvBench~\citep{huang2026rat,eliseeva2025envbench}.
We compare Graph2Env against static dependency extraction, specialized environment-construction systems, and general-purpose coding agents. Graph2Env achieves an  Environment Build Success Rate (EBSR) of 81.0\%, compared
with 71.5\% for Claude Code, the strongest general-purpose coding
agent evaluated, and 52.5\% for SetupX, the strongest specialized baseline on this
metric. Graph2Env further achieves an Environment Setup Success Rate (ESSR)
of 59.3\%, compared with 50.3\% for the strongest baseline.

In summary, this work makes the following contributions:
\begin{itemize}
    \item We introduce DepGraph, which supplies a dependency-ordered plan before execution and
    absorbs requirement-level information discovered during it.
\item We introduce Graph2Env, which externalizes the current state of repository environment construction into a typed dependency graph spanning heterogeneous requirements. DepGraph is continuously refined using execution evidence and operationalized by compiling its requirements and dependencies into a replayable construction procedure.
\item We evaluate Graph2Env against both specialized environment-construction systems and strong general-purpose coding agents, demonstrating substantial improvements in environment construction and downstream execution.
\end{itemize}

\section{Related Work}
\label{sec:related-work}

\subsection{Automated Repository Environment Construction}

Repository environment setup has traditionally been performed manually by developers following installation instructions. 
Early attempts at automation used static methods, deriving
dependencies from source analysis and knowledge bases of existing packages and
Dockerfiles~\citep{bndr2015pipreqs,horton2019dockerizeme,ye2021dockergen}. These methods
produce a requirement list before any command runs, but they compute it once and cannot
capture requirements that appear only when installation or execution fails. LLM-based agents
make setup interactive. The agents can inspect the repository, run commands in a container,
diagnose the resulting failures, and repair the environment~\citep{bouzenia2025executionagent,milliken2025installamatic,li2026repolaunch}. Recent
methods split this work across specialized roles or reuse experience from earlier
attempts~\citep{guo2025swefactory,wei2026sag,guo2026evoconfig}. These approaches organize the setup process in different ways, but the
construction state of the repository being configured remains implicit in the
interaction trajectory.
Among the works most relevant to ours, Repo2Run targets Python repositories, repairing
installation and test failures and writing the successful commands into a Dockerfile at the
end~\citep{hu2025repo2run}. RAT spans many language ecosystems and plans configuration in
advance, but over configuration modes and operations rather than over the requirements a
repository must satisfy~\citep{huang2026rat}. SetupX goes further and transfers experience from
earlier setup failures into later repairs~\citep{zhou2026setupx}. In all three, what a repository requires and what already holds
stay implicit in the trajectory. Graph2Env makes both explicit. DepGraph records the
requirements relevant to the designated target, their dependency and mapping relations, and
their states. The build script is compiled from that structure instead of assembled from a
successful interaction.

\subsection{Structured State, Memory, and Planning for {LLM} Agents}

Long-horizon agents often preserve information beyond the immediate action
through external memory and reflection. The Generative Agents framework~\citep{park2023generativeagents}
stores natural-language experiences, synthesizes reflections, and retrieves
relevant memories for later planning, while
Reflexion~\citep{shinn2023reflexion} converts task feedback into reflective
text retained for subsequent trials. More structured approaches organize
information for future decisions as cognitive maps or graph-based trajectory
abstractions: MAP~\citep{liu2026map}, a Map-then-Act paradigm, constructs a task-specific
cognitive map before execution, and Trainable Graph
Memory~\citep{xia2025graphmemory} abstracts trajectories into state-machine
decision paths and cross-task strategic memory. These approaches retain, retrieve, or organize task-relevant information to
support subsequent decisions, rather than explicitly modeling the current state of repository environment construction.

Graph2Env shares the broader goal of externalizing task-relevant information for subsequent decisions, but DepGraph is neither generic episodic memory nor learned cross-task strategy memory. It represents the active state of a specific repository-environment construction task: target-relevant requirements, their dependency relations, and evidence-derived requirement states. The interaction trajectory remains available as execution evidence, while DepGraph is refined from that evidence to support subsequent construction and repair. Thus, it keeps actionable construction information explicit without excluding the original interaction evidence.

\section{Graph2Env: The Proposed Approach}
\label{sec:method}

\label{sec:method-overview}
\suppressfloats[t]

\begin{figure}[t]
  \centering
  \includegraphics[width=\textwidth]{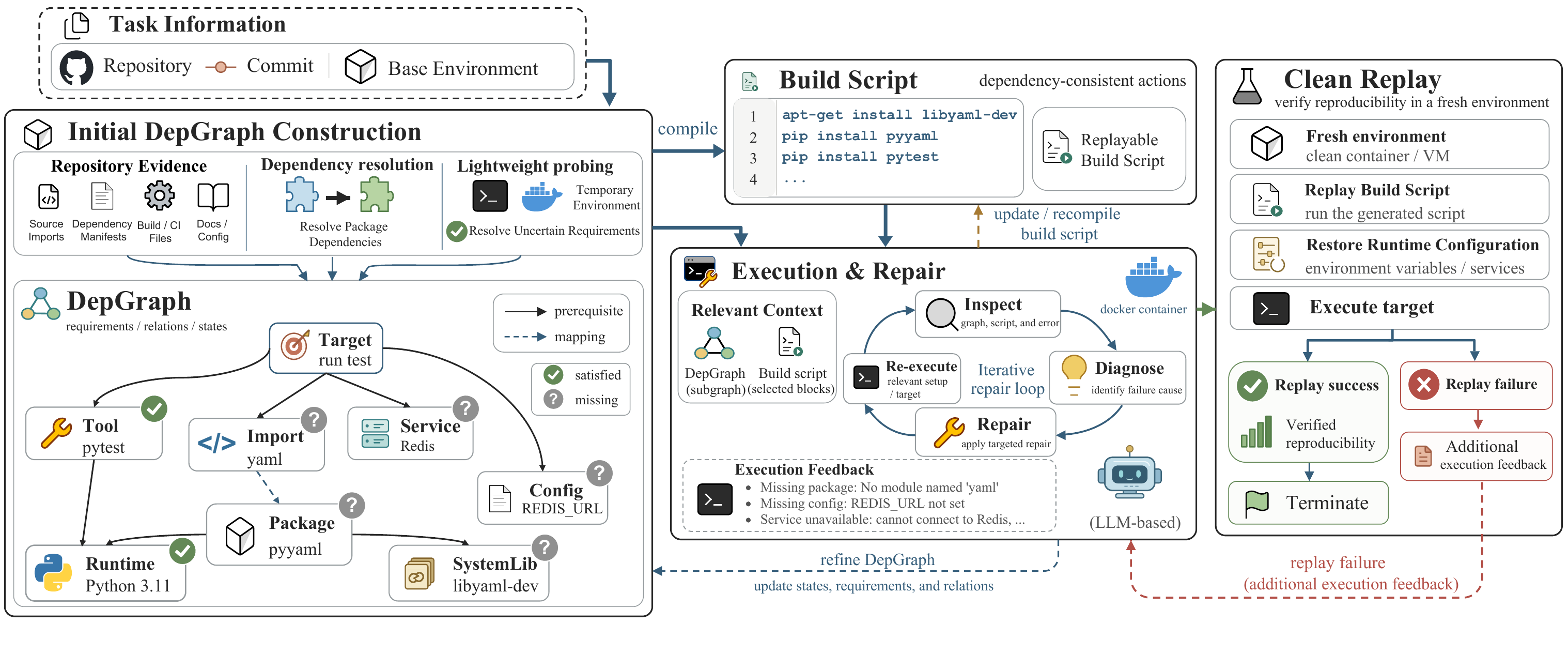}
  \caption{Overview of Graph2Env. Repository information is analyzed through dependency resolution and lightweight probing to construct DepGraph, which represents execution requirements, their relations, and satisfaction states. DepGraph is compiled into a replayable build script, and both provide structured context for execution and repair, while execution feedback refines the graph and updates the script. Once the designated target succeeds, Graph2Env replays the build script and restores runtime configuration in a fresh environment; replay success terminates construction, whereas replay failure returns additional feedback for further repair.}
  \label{fig:method-overview}
  \vspace{-12pt}
\end{figure}

Graph2Env constructs repository execution environments around a persistent DepGraph that explicitly represents the current construction state. We use \emph{execution target} to denote the repository operation that the constructed environment is intended to support, such as test collection or test execution. As shown in Figure~\ref{fig:method-overview}, Graph2Env first constructs an initial DepGraph from repository evidence, dependency resolution, and lightweight probing. It then compiles the graph into a dependency-consistent, replayable build script. During iterative construction, a DepGraph-guided agent diagnoses failures
observed while executing setup actions or the designated execution target and
uses the resulting feedback to refine both the graph and the persistent
construction state. Once the designated execution target succeeds in the working
environment, Graph2Env replays the resulting build script and
restores recorded runtime conditions in a fresh environment to
verify reproducibility. Appendix~\ref{app:case-study} provides
a concrete example of this workflow.

\subsection{DepGraph Construction}
\label{sec:depgraph-construction}

\textbf{DepGraph Representation.}
DepGraph is a typed directed graph
$G=(V,E)$,
which represents the environment requirements relevant to an execution target and their current satisfaction states. It contains a \textsc{Target} node representing the designated execution target, together with requirement nodes typed as \textsc{Import}, \textsc{Package}, \textsc{Runtime}, \textsc{Tool}, \textsc{SystemLib}, \textsc{Service}, or \textsc{Config}. \textsc{Import} and \textsc{Package} capture language- and package-level requirements; \textsc{Runtime}, \textsc{Tool}, and \textsc{SystemLib} capture the execution substrate; and \textsc{Service} and \textsc{Config} capture external services and runtime configuration. 
Edges encode prerequisite and mapping relationships. A prerequisite edge $u \rightarrow v$ indicates that satisfying requirement $u$ depends on prerequisite $v$ and is directed from the dependent to its prerequisite. A mapping relationship connects a source-level requirement to the installable requirement that satisfies it, such as mapping the import \texttt{docx} to the package \texttt{python-docx}. Each requirement node is annotated as \textsc{Missing} or \textsc{Satisfied} when available evidence supports either state; when the evidence is insufficient, its state remains unclassified. In this way, DepGraph maintains an explicit view of what the repository requires, how those requirements depend on one another, and which requirements have already been satisfied.

For example, Figure~\ref{fig:method-overview} shows a DepGraph for a repository whose
execution target is to run its tests. The graph records requirements such as
the \texttt{yaml} import, its mapped package \texttt{pyyaml}, the
\texttt{pytest} tool, the Python runtime, and runtime requirements such as the
Redis service and \texttt{REDIS\_URL} configuration. Their relations and
current states make explicit which requirements are already satisfied and
which remain to be resolved before the target can execute.

\textbf{Initial DepGraph Construction.}
Initial DepGraph construction is performed programmatically before the agent execution loop begins. Graph2Env first collects repository evidence from source-level imports, dependency manifests, build and CI files, service and configuration artifacts, and structured documentation. This evidence is converted into typed requirement nodes and known prerequisite and mapping relations. Graph2Env then performs dependency resolution to connect source-level requirements to installable packages and to identify additional package-level prerequisites under the relevant runtime and platform constraints. Specifically, Graph2Env uses Python package metadata to map source-level imports to installable distributions and uses \texttt{uv} to recursively resolve compatible direct and transitive package dependencies.

When repository evidence and dependency resolution are insufficient, Graph2Env performs lightweight probing in a temporary isolated environment to resolve uncertain requirements or environment assumptions. These probes are used only to acquire evidence for graph construction; their effects are discarded before agent execution begins. The resulting initial DepGraph therefore provides the agent with an explicit construction state before iterative execution and repair starts.

\subsection{DepGraph-Driven Compilation} \label{sec:depgraph-compilation} Given the current DepGraph, Graph2Env derives a setup-time action for each requirement whose resolved information determines an executable setup operation. For a requirement node $v$, the corresponding action is represented as an ordered command sequence \[ a(v)=\langle c_1,\ldots,c_m\rangle . \] Such actions may install language packages or system prerequisites, configure tools, or apply setup-time configuration; requirements without an executable setup operation contribute no action. For example, in Figure~\ref{fig:motivating-example}, the resolved package requirement \texttt{python-docx>=1.1.0} yields the installation action \texttt{python3 -m pip install 'python-docx>=1.1.0'}, illustrating how requirement information in DepGraph is translated into executable setup. Graph2Env then orders these actions according to the dependency relations in DepGraph. For an edge $u \rightarrow v$, where $u$ depends on prerequisite $v$, the action for $v$ precedes the action for $u$ whenever both have executable actions. Let $v_1,\ldots,v_k$ denote the requirements with executable actions in this dependency-consistent order. Graph2Env composes their actions into the replayable build script \[ P=a(v_1)\Vert a(v_2)\Vert\cdots\Vert a(v_k), \] where $\Vert$ denotes sequential concatenation. The resulting build script provides the initial executable setup for subsequent agent-guided construction. %

\subsection{DepGraph-Guided Execution and Repair}
\label{sec:execution-repair}

\textbf{DepGraph-Guided Diagnosis and Repair.}
Starting from the initial DepGraph and compiled build script, Graph2Env employs
an LLM agent that operates in a ReAct-style interaction loop~\citep{yao2023react} to iteratively
execute, diagnose, and repair the repository environment. At each step, the
agent observes the latest execution feedback and may selectively inspect
relevant portions of DepGraph and the corresponding build-script blocks. Based
on the available context, it selects and executes an action through the
Graph2Env action interface (Appendix~\ref{app:agent-actions}), and uses the resulting observation to
guide subsequent decisions. When an execution failure is observed, the agent combines the
execution error message with the relevant requirements, dependency relations, states,
and setup actions to diagnose the unresolved or incorrectly represented
requirement. 
It then applies a targeted repair, for example by installing a
missing dependency, modifying the relevant setup commands, or updating a
required runtime condition, and re-executes the relevant setup or execution
target to obtain new feedback.

\textbf{Execution-Grounded Graph Refinement.}
Execution feedback is used to refine Graph2Env's current representation of the repository requirements. Evidence about an existing requirement can update its state, while newly observed failures may reveal missing requirements, incorrect package mappings, or inaccurate dependency relations. Graph2Env updates DepGraph accordingly and recompiles the affected parts of the build script when the refinement changes executable setup information.

\textbf{Persistent Construction Artifacts.}
Repairs applied only to the current working environment may be lost when the environment is reconstructed, so Graph2Env records successful repairs in persistent construction artifacts rather than relying on transient environment state. Requirement and dependency information is retained in DepGraph, while setup-time repairs are incorporated into the build script and persistent runtime configuration is recorded alongside it when necessary. Once the designated target succeeds in the working environment, Graph2Env
performs a clean replay: it replays the build script and restores the recorded
runtime configuration in a fresh environment. If the replay successfully executes the designated target, Graph2Env considers
the construction complete and terminates the agent loop; otherwise, the replay
failure is returned as additional execution feedback and the repair process
continues until clean replay succeeds or the run reaches either the maximum
number of interaction steps or the time limit.

\vspace{-8pt}
\section{Experiments}
\label{sec:experiments}

\subsection{Benchmark}
\label{sec:benchmark}
Existing environment-construction benchmarks provide valuable repository
collections, but they do not directly provide the test-count reference required
by our evaluation protocol. In particular, neither
RATBench~\citep{huang2026rat} nor EnvBench~\citep{eliseeva2025envbench}
provides the number of tests that can be stably collected for each repository
in a certified working environment. Without such a fixed reference, a method
that fails to collect part of a repository's test suite may be evaluated against
a smaller test set induced by its own incomplete environment.

To address this limitation, we construct a new evaluation benchmark from
Python repositories drawn from RATBench and EnvBench. We focus on repositories
whose execution requirements can be addressed through software environment
construction and that require non-trivial setup and repair. As a lightweight
screening criterion, we prioritize repositories that the screening agent cannot
configure within 10 interaction steps. These repositories typically require
more than straightforward dependency installation and involve iterative
diagnosis and repair. Because establishing reliable test counts requires
multiple environment-construction attempts with a state-of-the-art coding
agent, the certification process is computationally and financially expensive;
we therefore limit the benchmark to 200 repositories.

For each selected repository, we independently establish a gold test count using
a state-of-the-art coding agent\footnote{Claude Code with Claude Opus 5.
Based on per-million-token pricing as of September 2026, Claude Opus 5 costs
\$5.00 / \$25.00 (input / output).}. The agent attempts to construct an environment in which
\texttt{pytest --collect-only} succeeds. An agent-constructed environment is
accepted only when integrity checks confirm that the repository has not been
modified and repeated test collection yields a stable result.
Across three independent attempts, we retain the largest successfully
certified test collection and use its size as the repository's gold test count.
This fixed reference prevents a method from being evaluated against a smaller
test set simply because its constructed environment fails to collect part of
the repository's tests. The complete certification protocol is provided in
Appendix~\ref{app:gold-manifest}.

\subsection{Evaluation Metrics}
\label{sec:metrics}

We adopt two metrics to evaluate the constructed environments: Environment
Build Success Rate (EBSR) from Repo2Run~\citep{hu2025repo2run} and
Environment Setup Success Rate (ESSR) from RAT~\citep{huang2026rat}.

\textbf{Environment Build Success Rate (EBSR).}
EBSR~\citep{hu2025repo2run} measures whether the constructed environment makes
the repository's test suite executable, regardless of whether individual tests
pass or fail. We report the percentage of repositories whose rebuilt image
returns exit code~0 from \texttt{pytest --collect-only}.

\textbf{Environment Setup Success Rate (ESSR).}
The EBSR evaluation checks if the test suite can be imported and discovered, but it does not execute the tests themselves. Some runtime requirements, such as external services, may only be exercised during test execution. ESSR~\citep{huang2026rat} evaluates this second aspect by executing
the test suite and measuring how much of the repository's recorded behaviour
can be reproduced. For repository $r$ with certified gold test count $D_r$, we
use the number of passing tests as the numerator,
$\mathrm{ESSR}_r=\min(\mathrm{passed}_r,D_r)/D_r$, and report the macro average.

\vspace{-8pt}
\subsection{Baselines and Implementation Details}
\label{sec:baselines}

We compare Graph2Env against six baselines spanning static dependency inference,
specialized environment-configuration agents, and general-purpose coding agents.
All agent systems, including Graph2Env, use DeepSeek-V4-Flash-0731\footnote{Based on per-million-token pricing via the official API,
\$0.22/\$0.66 (input/output).} as the LLM
backbone, with a limit of 100 LLM calls and two hours per
repository. 
All baselines are evaluated by rebuilding their Dockerfile or
equivalent build artifact in a fresh container to assess reproducibility.
Detailed configurations, prompts, and invocations are provided in Appendix~\ref{app:agent-settings}. 

\textbf{Static Dependency.}
\textbf{pipreqs}~\citep{bndr2015pipreqs} serves as a static dependency-inference
baseline. It scans repository source files to infer Python package requirements
from imports and generates a \texttt{requirements.txt} without any execution
feedback or iterative repair. We install the inferred requirements using a
fixed Dockerfile template and evaluate the resulting environment under the same
EBSR and ESSR protocol as the other methods.

\textbf{Environment Building Agents.}
\textbf{Repo2Run}~\citep{hu2025repo2run} runs under its own default
configuration; we supply the task, the model, and the step budget, and change
nothing else. RAT~\citep{huang2026rat} provides a specialized configuration toolset for
repository inspection, test collection, and environment repair. We evaluate its
\texttt{dockerfile} mode, which emits a recipe that rebuilds the configured
environment from a clean base. \textbf{SetupX}~\citep{zhou2026setupx} carries setup experience across
repositories. We initialize its experience store from the published pre-warmed
one and make it read-only, so the arm retrieves prior experience but accumulates
none and its repositories stay order-independent.

\textbf{Coding Agents.}
\textbf{SWE-agent}~\citep{yang2024sweagent} is a general-purpose coding agent that
explores a repository and edits files through a bash and file-editing interface.
We instruct each agent to inspect the repository, install what it needs, and
record the working setup as a Dockerfile. We also include \textbf{Claude Code}, a
state-of-the-art agent harness, evaluated under the same prompt as SWE-agent.
\\
We score every baseline from the artifact it produces, never from what it reports
about its own run. Four of the six configurations return a Dockerfile, which we
rebuild in a fresh container with no agent present; both metrics are computed
over the resulting image. 

\subsection{Experimental Results}
\label{sec:results}

\begin{table}[t]
\centering
\vspace{-8pt}
\caption{Environment construction success and agent token usage across baselines over the 200 repository benchmark. N/A marks a method that invokes no language model.}
\label{tab:baselines}
\setlength{\tabcolsep}{7.3pt}
\begin{tabular}{l l cccc}
\toprule
\rowcolor{gray!15}
Type & Method & EBSR & ESSR & \makecell{Tokens (K)} & Steps \\
\midrule
Static Dependency & pipreqs     & 10.0\% & 11.6\% & N/A & N/A \\
\midrule
\multirow{3}{*}{Environment Building Agents}
            & Repo2Run & 42.5\% & 34.7\% & 1128.6 & 37.5 \\
            & RAT      & 44.0\% & 40.4\% & 2066.2 & 62.0 \\
            & SetupX             & 52.5\% & 45.9\% & \textbf{913.6} & 85.0 \\
\midrule
\multirow{2}{*}{Coding agent}
            & SWE-agent   & 50.0\% & 35.6\% & 1894.7 & 71.7 \\
            & Claude Code & 71.5\% & 50.3\% & 1849.5 & 36.5 \\
\midrule
Ours        & \textbf{Graph2Env} & \textbf{81.0\%} & \textbf{59.3\%} & 1397.8 & \textbf{34.1} \\
\bottomrule
\end{tabular}
\vspace{-12pt}
\end{table}

Table~\ref{tab:baselines} reports EBSR, ESSR, mean tokens, and mean agent
steps for all methods. Across all baselines, Graph2Env
achieves the highest EBSR and ESSR, at 81.0\% and 59.3\%. It outperforms Claude
Code, the strongest baseline and a state-of-the-art coding agent, by 9.5 and 9.0
points, and SetupX, the strongest specialised repository environment-building agent, by
28.5 and 13.4 points. Notably, Claude Code outperforms all three specialized environment-building agents, while SWE-agent outperforms Repo2Run and RAT, suggesting that general-purpose coding agents can be competitive in repository environment construction even without being designed specifically for this task.

We also observe that Graph2Env completes environment construction with an average of 34.1 agent steps per repository, the lowest among the evaluated agent-based methods. However, it does not consume the fewest tokens: SetupX and Repo2Run use 913.6K and 1128.6K tokens on average, respectively, compared with 1397.8K for Graph2Env.
We attribute this to the DepGraph and compiled build script, which are
included in the prompt.

\begin{table}[t]
\centering
\caption{EBSR and ESSR (\%) by construction difficulty tier. $n$ is the number of repositories.}%
\label{tab:difficulty}
\setlength{\tabcolsep}{7pt}
\begin{tabular}{l r cc cc cc}
\toprule
\rowcolor{gray!15}
 & & \multicolumn{2}{c}{Graph2Env} & \multicolumn{2}{c}{Claude Code} & \multicolumn{2}{c}{Repo2Run} \\
\rowcolor{gray!15}
Tier & $n$ & EBSR & ESSR & EBSR & ESSR & EBSR & ESSR \\
\midrule
Easy     & 62 & \textbf{93.5} & \textbf{79.8} & 87.1 & 70.1 & 66.1 & 53.8 \\
Medium   & 86 & \textbf{82.6} & \textbf{57.5} & 75.6 & 48.3 & 40.7 & 30.5 \\
Hard     & 43 & \textbf{69.8} & \textbf{41.9} & 51.2 & 33.1 & 18.6 & 20.8 \\
\bottomrule
\end{tabular}
\end{table}

\textbf{Success by difficulty.}
We further group the 191 repositories with a certified, recorded labelling session by construction difficulty, measured as
the number of steps the labelling agent needed to establish each repository's total
test count: Easy ($\leq 15$), Medium ($16$--$45$), and Hard ($\geq 46$).
Graph2Env's EBSR lead over Claude Code is 6.4 points on Easy repositories
(93.5\% vs.\ 87.1\%) and 7.0 points on the Medium tier, and widens to
18.6 on the Hard tier (Table~\ref{tab:difficulty}). The ESSR margin stays
between 8.8 and 9.7 points across tiers. On the hardest repositories, Graph2Env
builds an environment far more often, but passing the full test suite remains
difficult for every system. Harder repositories require more system
libraries, native build headers, and runtime configuration, resolved over
longer repair sequences. DepGraph records which of these requirements are
satisfied, and the build script keeps every repair, rather than leaving both
to the agent's interaction history.
The results are consistent with our motivation: as construction requires longer repair sequences and involves more requirements, explicitly maintaining construction state becomes increasingly valuable relative to relying on interaction history alone.

\begin{wraptable}{r}{0.46\textwidth}
\vspace{-24pt}
\centering
\small
\setlength{\tabcolsep}{4pt}
\caption{SWE-agent with and without DepGraph as context, on all 200 repositories.}
\label{tab:ablation}
\begin{tabular}{lccc}
\toprule
\rowcolor{gray!15}
Configuration & EBSR & ESSR \\
\midrule
SWE-agent                  & 50.0\%          & 35.6\%         \\
\quad + graph context      & \textbf{62.0\%}\,{\scriptsize\color{green!50!black}$\uparrow$12.0}
                           & \textbf{42.4\%}\,{\scriptsize\color{green!50!black}$\uparrow$6.8}
                            \\
\bottomrule
\end{tabular}
\vspace{-8pt}
\end{wraptable}
\textbf{Does DepGraph help other agents?}
DepGraph drives every stage of Graph2Env---planning, repair, and script
compilation---so removing it leaves no working system to ablate against.
Instead, we ask whether the information DepGraph captures is useful to an
agent that was not designed around it. To evaluate it, we run SWE-agent on all 200
repositories with the same configuration as the baseline, but append
Graph2Env's initial DepGraph and the setup script compiled from it to its
prompt; both are produced before Graph2Env's repair loop. %

As shown in Table~\ref{tab:ablation}, the added context raises SWE-agent's
EBSR from 50.0\% to 62.0\% and its ESSR from 35.6\% to 42.4\%. DepGraph therefore
carries useful setup information that a generic agent can utilize. However,
SWE-agent with this context still trails Graph2Env (81.0\% EBSR), suggesting that the usefulness of DepGraph lies not only in the information it
initially captures, but also in its continuous refinement with execution feedback.

\begin{table}[!htbp]
\centering
\footnotesize
\setlength{\tabcolsep}{3.5pt}
\begin{minipage}[t]{0.52\linewidth}
\centering
\caption{Environment-building failure per agent}
\vspace{2pt}
\label{tab:failure-analysis}
\setlength{\tabcolsep}{6pt}
\begin{tabular}{l rrr}
\toprule
\rowcolor{gray!15}
Failure & G2E & CC & R2R \\
\midrule
No Dockerfile produced &    &    &    \\
\quad Time limit       & 10 & 0  & 16 \\
\quad Turn limit       & 6  & 1  & 0  \\
\quad Agent stopped    & 2  & 6  & 0  \\
\midrule
Image build fails      & 1  & 22 & 33 \\
Test collection fails  & 17 & 27 & 61 \\
Others                 & 2  & 1  & 5  \\
\midrule
\textbf{Total}         & \textbf{38} & \textbf{57} & \textbf{115} \\
\bottomrule
\end{tabular}
\\[2pt]{\footnotesize G2E: Graph2Env, CC: Claude Code, R2R: Repo2Run.}
\end{minipage}
\hfill
\begin{minipage}[t]{0.42\linewidth}
\centering
\makeatletter\def\@captype{figure}\makeatother
\caption{Failing tests by error type, in environments built by Graph2Env.}
\label{fig:error-types}
\includegraphics[width=\linewidth]{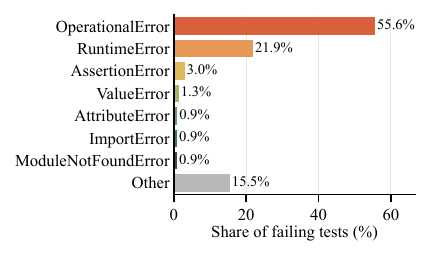}
\vspace{-12pt}
\end{minipage}

\vspace{-12pt} %
\end{table}

\subsection{Discussion}

\textbf{Failure analysis.} Although Graph2Env achieves the highest EBSR, it still fails on 38 of the 200 repositories (Table~\ref{tab:baselines}). We analyse failures at two levels: why a system fails to construct a working environment, that is, its rebuilt image does not pass test collection (Table~\ref{tab:failure-analysis}); and, within environments that do pass, why individual tests still fail at runtime (Figure~\ref{fig:error-types}).

Graph2Env's failures fall into two groups. In 16 of its 38 failures, the run exhausts its time or turn budget before exporting a Dockerfile, mostly on machine-learning and multi-service repositories whose packages are large and slow to install. The other 17 are collection failures, most of which pass Graph2Env's own clean rebuild but fail under the shared evaluator, which collects the full repository tree rather than the test paths each repository declares; these reflect a mismatch between the two collection procedures rather than an environment that cannot load its tests. Only one of Graph2Env's Dockerfiles fails to build, against 22 for Claude Code and 33 for Repo2Run, because Graph2Env accepts an environment only after replaying its build script and re-running the target in a fresh container. Claude Code instead accepts once collection succeeds in the container it has been modifying, which its Dockerfile may not reproduce, and Repo2Run exports a Dockerfile even when it exhausts its turn budget. The replay also explains the time limits: each replay repeats the full installation, which is costly on heavy repositories.

Figure~\ref{fig:error-types} shows the error types of failing tests in environments that Graph2Env successfully built. \texttt{OperationalError} (55.6\%) mainly occurs when a required service is not running, for example a PostgreSQL server that was never started, and \texttt{RuntimeError} (21.9\%) when runtime configuration such as an environment variable is missing. Both requirements surface only when tests execute: \texttt{pytest --collect-only} imports test modules without starting fixtures or services, so an environment can pass collection while lacking them.
Since most remaining failures stem from services and configurations that only matter at runtime, future work should explore more effective fault localization techniques, to help agents locate and fix runtime failures based on multimodal data such as error messages, logs, and agent trajectories.

\textbf{Generalizability}. Our evaluation focuses on Python repositories and uses a single LLM backbone, which may limit the direct generalizability of the reported results. However, the core design of Graph2Env is not specific to Python or a particular LLM. DepGraph represents environment requirements through general abstractions (such as runtimes, packages, and libraries)  rather than language-specific assumptions. Extending Graph2Env to other programming ecosystems would primarily require ecosystem-specific mechanisms for extracting repository evidence, resolving dependencies, and compiling requirements into executable setup actions. Similarly, the approach is conceptually compatible with different LLM backbones and agent harnesses. Indeed, the improvement obtained by providing DepGraph context to SWE-agent suggests that the representation can benefit an agent other than the one used by Graph2Env. %

\FloatBarrier
\section{Conclusion}

We have presented Graph2Env, a stateful approach to repository environment
construction that explicitly maintains %
environment requirements and
their dependencies through DepGraph. Experiments on a benchmark of 200 Python repositories drawn from RATBench
and EnvBench show that Graph2Env improves both environment build success and downstream test execution over specialized environment-construction
systems and general-purpose coding agents. These results suggest that explicitly
maintaining construction state provides a more reliable basis for iterative
environment construction than relying on the interaction trajectory alone. To facilitate replication, our source code and experimental data are
available at \url{https://anonymous.4open.science/r/Graph2Env}.

\subsection*{AI use statement}

Generative AI tools were used to assist with the implementation of research
code, the analysis and interpretation of experimental results, and language
editing of the manuscript. The research problem, core ideas, and method design
were developed by the authors. AI-assisted code was reviewed and tested by the
authors, and AI-assisted analyses were checked against the underlying
experimental results and logs. All AI-assisted manuscript edits were reviewed
and revised by the authors. LLMs and coding agents are also used as components
of the proposed method and evaluation procedure; these uses are described
separately in the corresponding method and experimental sections.

\subsection*{Reproducibility statement}

The anonymized source code for the core implementation of Graph2Env is
available at \url{https://anonymous.4open.science/r/Graph2Env}. Details of the evaluation set
construction and gold test count certification are provided in Appendix~\ref{app:dataset},
while the experimental settings and baseline configurations are documented in Appendix~\ref{app:agent-settings}.

\bibliography{references}
\bibliographystyle{arxiv_style}

\appendix
\section{Agent Action Interface}
\label{app:agent-actions}

This appendix summarizes the complete action interface exposed to the
Graph2Env Agent during environment construction and repair.

\paragraph{Selective Inspection.}
At any step, the Agent may request complete, filtered, or selected views of
DepGraph, and may inspect relevant setup blocks or bounded prior observations
on demand. These interfaces expose the construction state and execution
evidence pertinent to a decision without requiring the complete graph, build
script, or interaction history to be injected into every context.

{\footnotesize
\setlength{\tabcolsep}{2pt}
\setlength{\extrarowheight}{2pt}
\renewcommand{\arraystretch}{1.16}
\begin{longtable}{>{\raggedright\arraybackslash}p{0.17\textwidth}%
                         >{\raggedright\arraybackslash}p{0.27\textwidth}%
                         >{\raggedright\arraybackslash}p{0.50\textwidth}}
\caption{Agent-visible actions in the Graph2Env interface.}
\label{tab:agent-actions}\\
\toprule
\rowcolor{gray!15}
Category & Action & Description \\
\midrule
\endfirsthead
\caption[]{Agent-visible actions in the Graph2Env ReAct interface (continued).}\\
\toprule
\rowcolor{gray!15}
Category & Action & Description \\
\midrule
\endhead

Inspection & \texttt{inspect\_graph} & Returns the complete, filtered, or selected DepGraph view. \\
Inspection & \texttt{inspect\_setup} & Returns the complete build script or a stable block for a specified node. \\
Inspection & \texttt{inspect\_observation} & Returns a bounded exact slice of a prior raw execution observation. \\
\addlinespace[3pt]

Execution & \texttt{run\_shell} & Runs an Agent-provided shell command in the selected construction container. \\
Execution & \texttt{run\_block} & Runs the setup commands associated with one DepGraph node. \\
Execution & \texttt{run\_setup} & Runs the current complete build script. \\
Execution & \texttt{run\_collect} & Restores current runtime conditions and collects tests. \\
Execution & \texttt{run\_test} & Restores current runtime conditions and runs tests. \\
\addlinespace[3pt]

DepGraph update & \texttt{update\_graph} & Updates known requirements or adds newly discovered requirements and relations. \\
\addlinespace[3pt]

Build-script update & \texttt{recompile\_setup} & Regenerates the build script from the current DepGraph. \\
Build-script update & \texttt{update\_setup} & Appends, replaces, or updates a stable block in the build script. \\
\addlinespace[3pt]

Runtime update & \texttt{update\_runtime} & Updates declared services, environment variables, and runtime capabilities. \\
\addlinespace[3pt]

Candidate management & \texttt{create\_candidate} & Creates an isolated repair branch from the Active construction state. \\
Candidate management & \texttt{validate\_candidate} & Runs the build script and/or target in the current candidate. \\
Candidate management & \texttt{promote\_candidate} & Replaces the Active state with the current candidate state. \\
Candidate management & \texttt{abort\_candidate} & Discards the current candidate repair branch. \\
\addlinespace[3pt]

Verification & \texttt{run\_clean\_replay} & Replays the current build script, restores recorded runtime configuration, and executes the designated target in a fresh environment. \\
\addlinespace[3pt]

Control & \texttt{action\_batch} & Executes a sequence of actions and stops at the first failed sub-action. \\
Control & \texttt{finish} & Ends the Agent loop and records its completion reason. \\
\bottomrule
\end{longtable}
}

\section{DepGraph and Compilation Details}
\label{app:depgraph-details}

\paragraph{From Repository Evidence to Requirements.}
The evidence sources described in Section~\ref{sec:depgraph-construction}
provide complementary information about the environment required by the
execution target. Source-level imports identify modules used by the
repository, whereas dependency manifests specify installable packages and
their version constraints. Build and CI files provide evidence about
runtimes, tools, and setup procedures; service and configuration artifacts
describe conditions required during execution. Structured documentation
provides additional evidence where these conditions are not fully expressed
in executable configuration. Graph2Env represents these observations as typed requirement nodes,
with prerequisite and mapping relations connecting the corresponding
requirements. In particular, an imported module and an
installable package remain distinct requirements: observing an import
establishes a language-level need, but does not by itself establish which
distribution provides it or whether that distribution is available in the
environment.

\paragraph{Import Mapping and Dependency Resolution.}
Import-to-distribution mapping connects source-level requirements to
packages that can satisfy them. This distinction matters when module and
distribution names differ, as in the mapping from \texttt{docx} to
\texttt{python-docx}. Python package metadata provides evidence for these
associations; for installed distributions,
\texttt{importlib.metadata.packages\_distributions()} exposes the
distributions associated with each top-level module. Package resolution
then determines compatible versions and expands package-level
prerequisites using \texttt{uv}, subject to the selected runtime and
platform constraints. Direct requirements originate from repository
evidence, while transitive requirements arise from the dependencies of
the resolved packages. Both are represented in DepGraph, allowing a
requirement observed in repository code to be related to prerequisites
that are not explicitly named there. Resolution identifies a candidate
package configuration; whether it supports execution must still be
established through observations from the environment.

\paragraph{Evidence from Lightweight Probing.}
Probing supplements repository evidence when a requirement cannot be
determined from declarations or dependency resolution alone. In a
temporary isolated environment, Graph2Env checks relevant runtime and
platform properties and observes whether selected packages can be
installed or imported. These observations can expose prerequisites that
package declarations do not fully describe, including tools needed during
installation and native libraries needed when a module is loaded. The
resulting evidence refines the requirements and relations recorded in
DepGraph. These observations describe the temporary probe environment and serve
as evidence for subsequent construction. The environment and its
modifications are discarded before agent execution; only the resulting
requirement information and observations are retained.

\paragraph{Compilation Details.}
For each requirement node with a known executable setup procedure,
$a(v)$ denotes its ordered command sequence. Graph2Env derives these
commands by instantiating predefined installation and configuration
rules with information obtained from repository evidence and dependency
resolution, such as package names, version constraints, and tool
requirements. Applicable setup commands extracted from repository build
and CI files can also supply steps in the sequence. During subsequent
repair, the agent may revise or extend these commands in response to
execution feedback. Such procedures may install packages, set up tools
or system prerequisites, or apply setup-time configuration; not every
requirement node has an executable action.

Graph2Env orders the actions using prerequisite relations, with an edge
$u \rightarrow v$ indicating that $a(v)$ should precede $a(u)$ when both
requirements have executable actions. The ordering procedure prioritizes actions whose prerequisite actions have already been scheduled. The resulting actions are
concatenated into the replayable build script, preserving the command
order within each action. Persistent runtime conditions, such as
environment variables and service startup requirements, are maintained
separately from the build script and restored when the designated target
is executed.

\section{An Example of Graph2Env}
\label{app:case-study}

\begin{figure}[!t]
    \centering
    \includegraphics[
        width=\linewidth
    ]{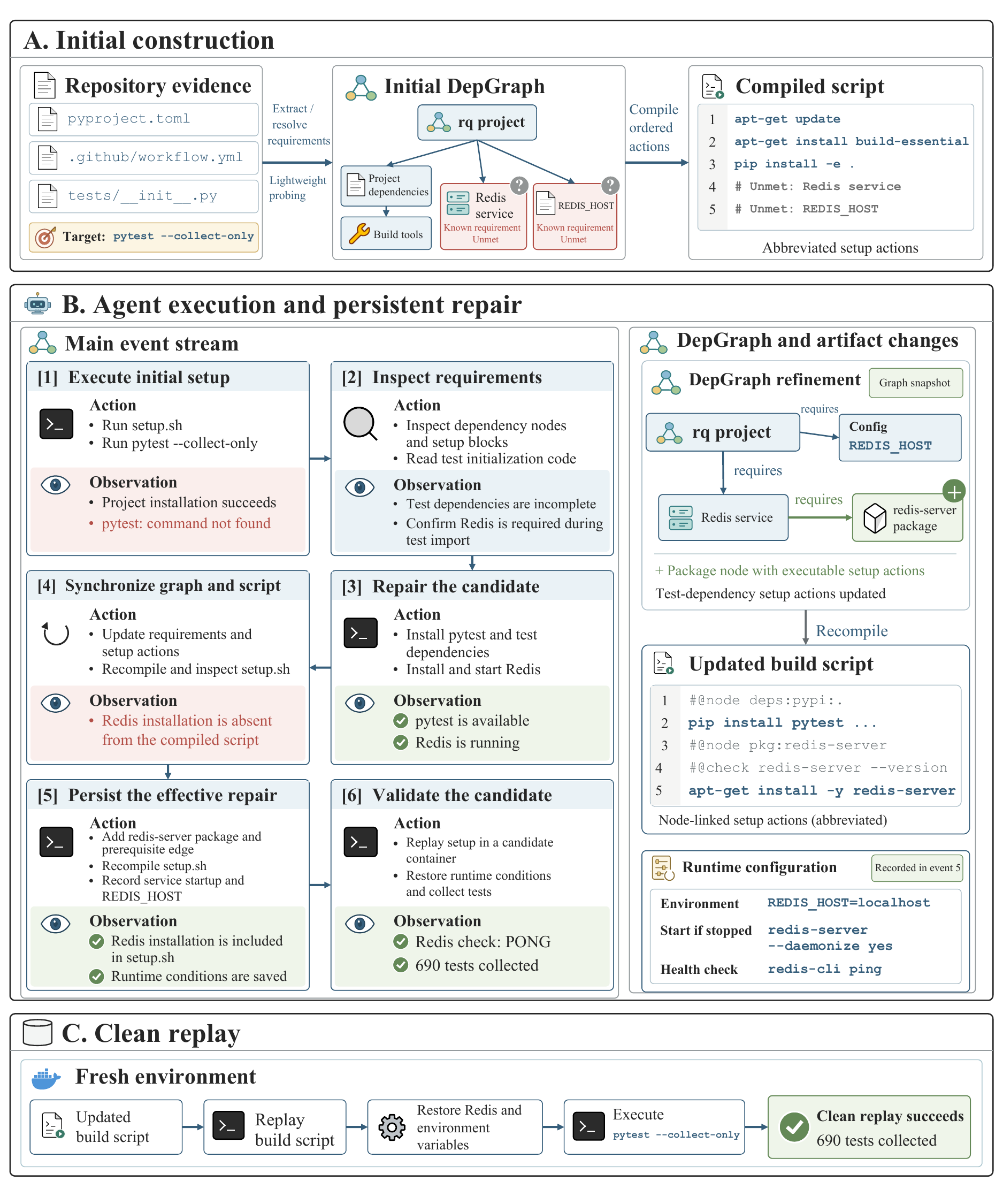}
    \caption{
        An environment construction example for \texttt{rq/rq}.
        (A) Initial requirement discovery and DepGraph-driven compilation.
        (B) Execution-guided repair updates the graph, build script,
        and runtime configuration.
        (C) Clean replay reconstructs the environment and successfully
        collects 690 tests.
    }
    \label{fig:rq-case-study}
\end{figure}

Figure~\ref{fig:rq-case-study} shows how Graph2Env constructs an
environment for \texttt{rq/rq} at revision \texttt{eacec8ff9109} to run
\texttt{pytest --collect-only}: requirement discovery, DepGraph
compilation, execution-guided repair, and clean replay.

\paragraph{Initial construction (A).}
Panel~A illustrates how Graph2Env gathers evidence from repository
metadata, build and workflow configurations, and test code to
identify environment requirements. Graph2Env constructs the initial DepGraph through requirement
discovery and dependency resolution, then compiles its available
setup actions into a build script. Redis and \texttt{REDIS\_HOST}
are already recorded as requirements, although their setup
procedures remain incomplete.

\paragraph{Execution and persistent repair (B).}
Execution exposes missing test dependencies. The agent inspects
the graph and test code, installs the required dependencies,
and starts Redis. Inspection of the recompiled script reveals
that Redis installation has not yet been preserved. The agent
therefore adds a \texttt{redis-server} package node and its
prerequisite relation, recompiles the script, and records the
runtime configuration. Validation then successfully collects
690 tests.

\paragraph{Clean replay (C).}
Graph2Env replays the updated script in a fresh environment,
restores the recorded runtime configuration, and successfully
collects 690 tests again. This verifies that the persisted
construction artifacts reproduce an environment supporting
the designated collection target.

\section{Dataset Construction and Analysis}
\label{app:dataset}

\subsection{Corpus Composition and Scale}
\label{app:scale}

We report summary statistics of the benchmark to characterize the diversity of the selected
repositories. The benchmark comprises 200 Python repositories drawn from
RATBench~\citep{huang2026rat} and EnvBench~\citep{eliseeva2025envbench} under the selection
criteria of Section~\ref{sec:benchmark}: 100 RATBench-sampled and 100 EnvBench-sampled
repositories, each pinned to a fixed commit. No repository is sampled from both benchmarks.

\paragraph{Language scope.}
All repositories are Python projects.
Python accounts for a median 99\% of a repository's source lines, but 54 of the 200
repositories contain more than 10\% non-Python code (C/C++, Cython, JavaScript, or shell),
typically native extensions or bundled front ends that must be compiled or resolved
before tests can import.

\begin{figure}[!hbp]
  \centering
  \includegraphics[width=\columnwidth]{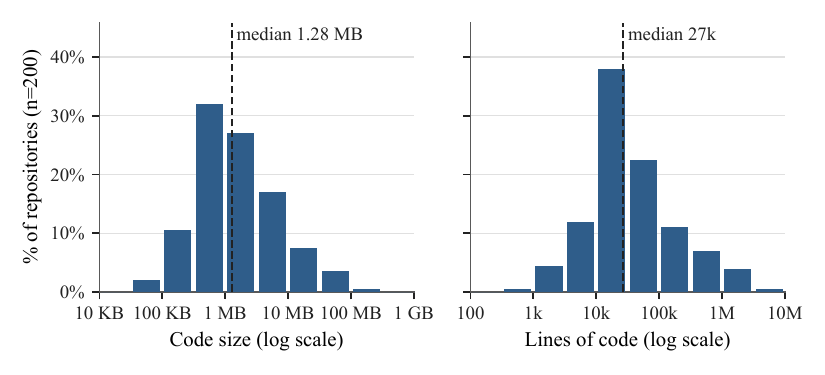}
  \caption{Scale of the 200 benchmark repositories (RATBench- and EnvBench-sampled, pooled). Left: code
  size (\texttt{code\_bytes}, GitHub's per-language byte count; median 1.28\,MB, IQR
  589\,KB--3.62\,MB). Right: non-blank lines of source code at the pinned commit, excluding
  vendored and minified files (median 27k, IQR 13k--88k). The two measures agree closely
  (Spearman $\rho = 0.89$).}
  \label{fig:dataset-scale}
\end{figure}

\paragraph{Repository size distribution.}
We measure size in two ways: code bytes, as reported by GitHub's per-language statistics
(excluding non-code assets), and lines of code (LoC), counted as non-blank lines in source
files at the pinned commit, excluding vendored and minified files. Figure~\ref{fig:dataset-scale}
shows both distributions. Repositories span more than three orders of magnitude, from
961 to 4.7M LoC, with a median of 1.28\,MB (IQR 589\,KB--3.62\,MB) and 27k LoC (IQR 13k--88k).
The two measures agree closely (Spearman $\rho = 0.89$). Using the size tiers of
RATBench~\citep{huang2026rat}, 35 repositories are Small ($<$500\,KB), 125 Medium
(500\,KB--5\,MB), and 40 Large ($>$5\,MB). The two samples differ in scale:
RATBench-sampled repositories skew larger (median 2.0\,MB; 33 of the 40 Large), while
EnvBench-sampled repositories account for most of the Small tier (23 of 35). Size matters
for environment construction because larger repositories reach further outside the Python
package manager, into system libraries and native toolchains.

\begin{wraptable}{r}{0.30\textwidth}
\vspace{-12pt}
\centering
\small
\caption{GitHub stars.}
\label{tab:stars}
\begin{tabular}{lr}
\toprule
\rowcolor{gray!15}
Stars & \% of repos \\
\midrule
$<$100 & 22\% \\
100--1k & 31\% \\
1k--10k & 36\% \\
$>$10k & 11\% \\
\bottomrule
\end{tabular}
\vspace{-8pt}
\end{wraptable}

\paragraph{Repository popularity distribution.}
Popularity, measured by GitHub stars, is long-tailed (range 12 to 155,569; median 786, mean
4,536). The corpus spans the full range (Table~\ref{tab:stars}): 44 repositories have fewer
than 100 stars, 62 have 100--1k, 72 have 1k--10k, and 22 exceed 10k. The benchmark is therefore not confined to
either little-known repositories, which are often under-documented, or widely used projects,
which often ship mature CI and container configurations.

\subsection{Application Domains}
\label{app:domains}

\begin{table}[t]
\centering
\small
\caption{Application domains of the 200 benchmark repositories. Each repository is assigned one
primary domain from its PyPI \texttt{Topic} classifiers where they give a unique majority, and
otherwise by an LLM from its description, GitHub topics, and README.}
\label{tab:domains}
\begin{tabular}{lrrrrrrrr}
\toprule
\rowcolor{gray!15}
 & Data/ML & Dev tools & Scientific & Web & Infra/DB & Security & Other & Total \\
\midrule
Repos & 46 & 38 & 35 & 32 & 30 & 4 & 15 & 200 \\
\bottomrule
\end{tabular}
\end{table}

\paragraph{Domain distribution.}
We assign each repository one primary application domain, defined by what the software is
for rather than the tools it is built with. Table~\ref{tab:domains} shows the distribution.
No single domain accounts for more than 23\% of the corpus: data and machine learning (46),
developer tools (38), scientific computing (35), web applications and APIs (32), and
infrastructure and database clients (30) are each well represented, with 19 repositories in
security or other domains. The two samples are complementary. EnvBench-sampled repositories
account for most of the scientific (28 of 35) and developer-tool (24 of 38) repositories,
while RATBench-sampled repositories account for most of the data/ML (30 of 46) and
infrastructure (24 of 30) repositories. Combining the two benchmarks therefore yields a
more balanced domain mix than either provides alone.

\subsection{Gold Test Count Certification}
\label{app:gold-manifest}

We describe how the gold test count of Section~\ref{sec:benchmark} is established for each
repository. The count is fixed before any evaluated system is run and does not depend on
any of them.

\paragraph{Acceptance criteria.}
An environment is certified only if it satisfies three conditions. First,
\texttt{pytest -{}-collect-only} exits with status~0 on two independent executions in the
built image. Second, both executions return the identical set of test node IDs, which
rejects environments whose collection depends on execution order or state. Third, every
tracked file in the repository has the same hash before the build and inside the built
image, so the count cannot be raised by adding, removing, or editing tests or their
configuration. The gold test count is the size of the certified node-ID set.

\paragraph{Procedure and budget.}
For each repository at its pinned commit, a coding agent is given a single objective: produce an
environment that meets the acceptance criteria and collects as many tests as possible. The agent may
edit only the Dockerfile, starting from a seed that copies the source tree onto a
\texttt{python:3.11-slim} base image and attempts an editable install; a session that did not reach
certification could be continued from the Dockerfile it left behind. Each session is limited to one
hour of wall-clock time with no turn limit, and when several sessions were run, the accepted one
collecting the most tests is retained. The retained sessions took a median of 7.0 minutes and
\$1.05 per repository.

\subsection{Certification Agent Configuration}
\label{app:gold-prompt}

The certifying agent is driven as a headless CLI in a container holding the repository
at its pinned revision. Each repository receives three independent attempts; the
Dockerfile is reset to the seed between attempts, and there is no turn limit.
\begin{lstlisting}[style=promptstyle]
# model: claude-opus-5
# attempts: 3 per repository, Dockerfile reset to seed between attempts
# budget:   no turn limit; 3600 s wall-clock per attempt

claude -p "<task prompt below>" \
  --dangerously-skip-permissions \
  --model <model> \
  --output-format stream-json --verbose
\end{lstlisting}

The agent starts from the following seed Dockerfile, so every package, system library,
and environment variable in a certified Dockerfile was added by the agent.

\begin{lstlisting}[style=promptstyle]
FROM <base>
WORKDIR /src
COPY . /src
RUN pip install --no-cache-dir -e . || pip install --no-cache-dir . || true
\end{lstlisting}

The task prompt is reproduced verbatim.

\begin{lstlisting}[style=promptstyle,literate={—}{{\textemdash}}1]
You are configuring a reproducible test-COLLECTION environment for a Python repository. Your ONLY editable file is `Dockerfile` in this directory. Your goal: make the repository's full pytest suite collect cleanly and maximally.

SUCCESS CRITERION. Run `./verify`. It builds your Dockerfile from scratch, runs `pytest --collect-only` inside the image twice, and reports:
- `accepted: true` when collection is clean (no collection errors) and stable across both runs. REQUIRED.
- `collected=N` — number of tests collected. MAXIMIZE this.
- `import_skipped=[modules]` — modules pytest skipped at import time, usually a missing optional dependency hiding real tests.
You are done when `./verify` reports accepted AND `collected` is as high as it will go — i.e. `import_skipped` contains only modules that are genuinely optional or whose dependency truly cannot be installed. A clean collection that hides half the suite behind missing deps is a FAILURE, not a pass.

THE ONLY LEVER IS THE ENVIRONMENT. `pytest --collect-only` works by IMPORTING every test module, so collection fails or shrinks only because the environment is missing something: a collection error (ImportError/ModuleNotFoundError) means a dependency isn't installed; an import-skipped module means an optional dep behind importorskip(...) isn't installed. The fix is always to install the missing dependency in the Dockerfile — read the failing import in the traceback, find the PyPI or apt package that provides it, and add it.

SERVICES (databases, brokers, etc.). `pytest --collect-only` only imports modules — it never runs tests or fixtures — so tests needing a live service (Redis, Postgres, RabbitMQ, ...) to PASS still collect fine without that service running. Install the service's Python CLIENT LIBRARY (e.g. redis, psycopg2-binary, pika) when a module fails to import it, but do NOT try to start the actual database/broker: it isn't needed for collection, and collection runs with NO network access. The one exception is a module that opens a connection at import time (top-level code, not inside a test or fixture) — live services can't be provided during collection, so leave those in import_skipped, install the client library, and move on.

RULES.
- Edit ONLY the `Dockerfile`. Do NOT touch tests, conftest.py, pyproject.toml, pytest.ini, setup.cfg, tox.ini, or any source file. The harness restores all of these to their pinned originals and hash-checks them before certifying — any edit you make is reverted and rejected, so it cannot help you.
- Do NOT fake a clean collection by hiding tests: no --ignore/-k/-m/--deselect, no collect_ignore/norecursedirs, no deleting or emptying test files or narrowing paths. All rejected. The only path that works is installing dependencies.
- Do NOT install anything that randomizes collection (e.g. pytest-randomly); the node-ID set must be identical across both runs.
- Do NOT run test bodies — only collection matters; tests never need to pass, only to be importable.
- The Dockerfile must build cleanly from scratch; no reliance on host state.

SUGGESTED WORKFLOW. Run `./verify` -> read the first traceback -> install the missing import in the Dockerfile, preferring the repo's DECLARED test/dev groups first (`pip install -e .[test]`/`[dev]`/`[all]`, requirements-dev.txt, test-requirements.txt) and `apt-get install` for C-library imports -> re-run. Repeat until accepted and `import_skipped` is minimal. Declared dependency groups usually close most collection gaps at once.

ZERO COLLECTION ERRORS — THERE IS NO PARTIAL CREDIT. The gate requires pytest to exit 0 on BOTH runs. ONE un-importable test module rejects the ENTIRE repository, no matter how many thousands of tests collected around it. So never stop at the first traceback: after each `./verify`, enumerate EVERY distinct error and fix them all. A big repo routinely hides a dozen unrelated missing distributions behind the first one, and a run that fixes nine of ten scores exactly the same as one that fixes none.

HARD LIMITS OF THE COLLECTION SANDBOX. `./verify` collects inside a locked-down container: NO NETWORK (`--network none`), 2 CPUs, 4 GB RAM, 512 processes, all capabilities dropped. Design the Dockerfile around this:
- Everything a module needs AT IMPORT must be baked in at BUILD time. Nothing can be fetched during collection — no pip, no model weights, no fixture data, no NLTK/HuggingFace caches. Pre-fetch it in a RUN layer.
- Importing a heavyweight framework (torch, tensorflow, jax) can exhaust the 4 GB cap; the OOM surfaces as a confusing collection error or a killed process, not an ImportError. Prefer CPU-only wheels (`tensorflow-cpu`, torch's `+cpu` index) — they import in a fraction of the memory.
- A module that opens a socket at import time cannot be satisfied (no network). Install its client library anyway; that is often enough, because the connection is usually built lazily inside a fixture.

STABILITY — THE NODE-ID SET MUST BE IDENTICAL ACROSS THE TWO RUNS. If `./verify` reports the set unstable, something is nondeterministic: parametrize IDs derived from set/dict iteration, a timestamp, a uuid, a tempfile name, or filesystem glob order — or an installed plugin that generates tests dynamically. `ENV PYTHONHASHSEED=0` fixes the most common case (hash-order-dependent IDs). A repo can collect tens of thousands of tests cleanly and STILL be rejected on this clause alone, so if you see it, fix it — it is not cosmetic.
\end{lstlisting}

\section{Image Size}
\label{app:dockerfile-quality}

Beyond whether an environment builds, we record the size of each successfully
built image, since smaller images are cheaper to store, transfer, and start at
scale.

\begin{table}[t]
\centering
\caption{Built Dockerfiles and image size across baselines over the 200 repositories.
\# Built counts Dockerfiles that built successfully; Image (MB) is the mean over built images.}
\label{tab:dockerfile-quality}
\setlength{\tabcolsep}{7pt}
\begin{tabular}{l l cc}
\toprule
\rowcolor{gray!15}
Type & Method & \# Built & Image (MB) \\
\midrule
Static Dependency & pipreqs     & 111 & 1616.5 \\
\midrule
\multirow{3}{*}{Environment Building Agents}
            & Repo2Run & 147 & 4145.5 \\
            & RAT      & 159 & 1577.0 \\
            & SetupX   & 126 & 1595.5 \\
\midrule
\multirow{2}{*}{Coding agent}
            & SWE-agent   & 111 & 1309.4 \\
            & Claude Code & 171 & 983.6  \\
\midrule
Ours        & Graph2Env   & 180 & 3309.8 \\
\bottomrule
\end{tabular}
\end{table}

As shown in Table~\ref{tab:dockerfile-quality}, Graph2Env produces the most
buildable Dockerfiles (180, vs.\ 171 for the strongest baseline, Claude Code),
but its images are larger on average (3309.8 MB vs.\ 983.6 MB). Inspecting its
setup scripts, we find that Graph2Env installs a compiler toolchain in 139 of
200 repositories so that source builds succeed, and seldom clears the pip or
apt caches afterwards (7 and 1 repositories, respectively). Both choices
increase size without affecting whether the environment works, and could be
removed with a post-install cleanup step.

\section{Additional Experiment Results}
\label{app:additional-results}

\paragraph{Results by domain.}
Table~\ref{tab:results-by-domain} reports EBSR and ESSR by application domain. Graph2Env achieves
the highest EBSR and ESSR in five of the seven domains. Its largest EBSR margin is on web
repositories (96.9\% against 65.6\% for Claude Code), and its smallest are on scientific (2.9 points)
and infrastructure (3.3 points) repositories. Claude Code leads on the security and other domains,
which together contain 19 repositories. Data/ML repositories are among the most difficult for all
three methods.

\paragraph{Results by repository size.}
Table~\ref{tab:results-by-size} reports the same metrics by code size. Success decreases with size
for every method: Graph2Env's EBSR falls from 100.0\% on small repositories to 83.2\% on medium and
57.5\% on large ones. Graph2Env achieves the highest EBSR and ESSR in every size tier, with an EBSR
margin over Claude Code of 7.5 to 10.4 points.

\begin{figure}[!htbp]
\centering
\includegraphics[width=\columnwidth]{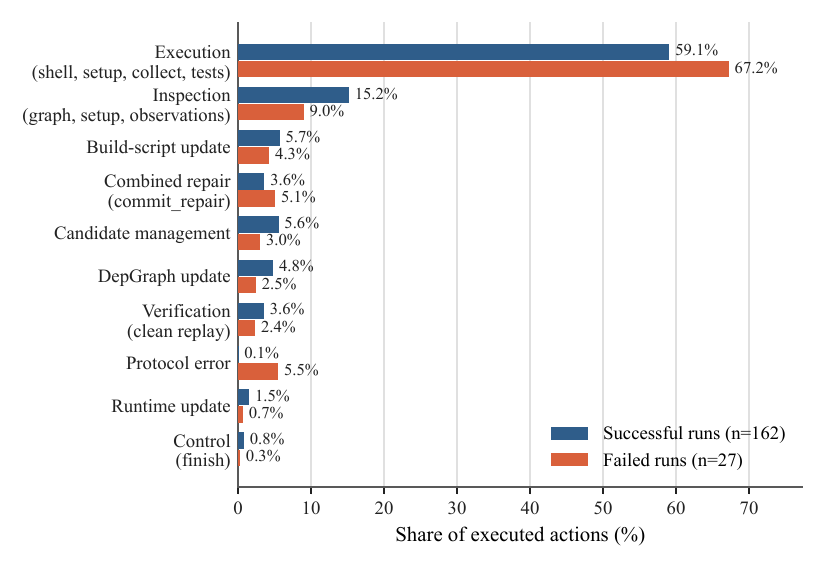}
\caption{Distribution of Graph2Env's executed actions by category, for successful (EBSR = 1, $n$=162)
and failed ($n$=27) runs. The remaining 11 failed runs (10 at the time limit, 1 crashed) were
terminated before their action trace was written and are excluded. Each bar is the
share of that group's actions. Actions issued inside an \texttt{action\_batch} are counted
individually; the batch itself is not. Categories follow Table~\ref{tab:agent-actions}.}
\label{fig:action-categories}
\end{figure}

\paragraph{Successful and failed runs.}
Figure~\ref{fig:action-categories} compares the action distribution of successful and failed runs.
Successful runs are shorter (median 27.5 actions against 89) and allocate more of their actions to
inspection (15.2\% against 9.0\%) and DepGraph updates (4.8\% against 2.5\%), whereas failed runs
spend more on execution (67.2\% against 59.1\%). Harder repositories may both fail and require
longer runs, so this pattern is correlational.

\paragraph{Use of construction state.}
Most shell activity is diagnostic: 81\% of \texttt{run\_shell} calls are declared read-only. Changes
to the environment are recorded in persistent artifacts, with more persistence actions than
state-changing shell commands. DepGraph is maintained throughout construction: it is updated in 164
of the 189 repositories with a recorded action trace (Figure~\ref{fig:action-categories}) and inspected in 136, at a cost of about 9\% of all actions.

\begin{table}[!htbp]
\centering
\small
\setlength{\tabcolsep}{2.2pt}
\caption{EBSR / ESSR (\%) by application domain. $n$ is the number of repositories in each domain.}
\label{tab:results-by-domain}
\begin{tabular}{lccccccc}
\toprule
\rowcolor{gray!15}
 & Data/ML & Dev tools & Scientific & Web & Infra/DB & Security & Other \\
\rowcolor{gray!15}
Method & ($n$=46) & ($n$=38) & ($n$=35) & ($n$=32) & ($n$=30) & ($n$=4) & ($n$=15) \\
\midrule
Repo2Run    & 23.9 / 22.0 & 60.5 / 58.5 & 45.7 / 39.2 & 43.8 / 33.1 & 43.3 / 26.6 & 50.0 / 24.9 & 40.0 / 24.8 \\
Claude Code & 60.9 / 40.0 & 73.7 / 71.5 & 85.7 / 56.5 & 65.6 / 40.6 & 66.7 / 41.9 & 75.0 / 32.4 & 86.7 / 55.4 \\
Graph2Env   & 69.6 / 46.7 & 89.5 / 82.0 & 88.6 / 62.7 & 96.9 / 58.8 & 70.0 / 53.4 & 75.0 / 27.6 & 66.7 / 54.1 \\
\bottomrule
\end{tabular}
\end{table}

\begin{table}[!htbp]
\centering
\small
\setlength{\tabcolsep}{4.5pt}
\caption{EBSR / ESSR (\%) by repository code size.}
\label{tab:results-by-size}
\begin{tabular}{lccc}
\toprule
\rowcolor{gray!15}
 & Small & Medium & Large \\
\rowcolor{gray!15}
Method & ($<$500\,KB, $n$=35) & (500\,KB--5\,MB, $n$=125) & ($>$5\,MB, $n$=40) \\
\midrule
Repo2Run    & 74.3 / 58.1  & 41.6 / 35.1 & 17.5 / 12.8 \\
Claude Code & 91.4 / 71.0  & 72.8 / 51.7 & 50.0 / 27.6 \\
Graph2Env   & 100.0 / 76.7 & 83.2 / 63.0 & 57.5 / 32.6 \\
\bottomrule
\end{tabular}
\end{table}

\section{Agent Settings}
\label{app:agent-settings}
This appendix gives the exact configuration of each baseline arm in Section~\ref{sec:baselines}.

\subsection{pipreqs}
\label{app:baseline-pipreqs}

pipreqs 0.4.13 scans the repository at its pinned commit and writes the
imported third-party packages to a requirements file. It uses no LLM and
receives no execution feedback. We use \texttt{no-pin} mode, which lists bare
package names rather than pinning them to the newest PyPI release at scan time,
so that reruns on the same commit produce the same file.

\begin{lstlisting}[style=promptstyle]
# pipreqs 0.4.13, run on the repository at its pinned commit
python -m pipreqs.pipreqs <repo> \
  --savepath requirements_pipreqs.txt \
  --force \
  --ignore .venv,venv,env \
  --mode no-pin \
  --encoding utf-8-sig
\end{lstlisting}

The requirements are installed with a fixed Dockerfile template. A requirement
that fails to install fails the build.

\begin{lstlisting}[style=promptstyle]
FROM python:3.10
WORKDIR /
RUN pip install pytest pytest-xdist
RUN git clone <repo_url>.git /testbed
RUN git -C /testbed fetch --depth 1 origin <commit> \
 && git -C /testbed checkout --detach <commit>
WORKDIR /testbed
COPY requirements_pipreqs.txt /requirements_pipreqs.txt
RUN pip install -r /requirements_pipreqs.txt
\end{lstlisting}

\subsection{Repo2Run}
\label{app:baseline-repo2run}

Repo2Run at commit \texttt{65042aa}, run under its own default configuration.
We supply the task, the model, and the turn budget; its prompts, tool
definitions, and control flow are upstream stock.

\begin{lstlisting}[style=promptstyle]
# repo2run @ 65042aa  --  github.com/bytedance/repo2run
python3 build_agent/main.py \
  --full_name <owner/repo> \
  --sha       <dataset base commit> \
  --root_path <run directory> \
  --num_turn  100 \
  --llm       deepseek/deepseek-v4-flash-0731
\end{lstlisting}

\subsection{SetupX}
\label{app:baseline-setupx}

SetupX at 85de355. SetupX produces no Dockerfile. It mutates a live container and records the
surviving commands to a history file, which we replay onto a pinned clone to
obtain a build artifact comparable to the other baselines. SetupX also maintains
an experience store that carries setup experience between repositories. We
initialize it from the pre-warmed store published in the official repository and
make it read-only for the evaluation, so the arm retrieves prior experience but
accumulates none across the benchmark run.

\begin{lstlisting}[style=promptstyle]
# SetupX @ 85de355
LLM_PROVIDER=openai
OPENAI_BASE_URL=https://api.deepseek.com/v1
OPENAI_MODEL=deepseek-v4-flash-0731
SETUPX_MAX_LLM_CALLS=100              # added; counts completions, not agent steps
DOCKER_BASE_IMAGE=python:3.10         # SetupX's own default
DOCKER_WORK_DIR=/workspace            # repository at /workspace/repo
SETUPX_DB_DSN=<dsn>                   # set: XPU experience store on; unset: off
XPU_ENABLED=1
XPU_VECTOR_ENABLED=1
XPU_READONLY=1                        # added; disables experience writeback
\end{lstlisting}

\subsection{RAT}
\label{app:baseline-rat}

RAT at commit \texttt{747adef}, run under its own default configuration with
\texttt{save\_mode} set to \texttt{dockerfile}, so that it emits a Dockerfile that
rebuilds the environment from a clean base.

\begin{lstlisting}[style=promptstyle]
# RAT @ 747adef  --  github.com/gemelom/RunAnyThing

python env_main.py \
  --full_name <owner/repo> \
  --root_path <run directory> \
  --num_turn  100 \
  --llm       deepseek/deepseek-v4-flash-0731 \
  --save_mode dockerfile

# upstream defaults, unchanged:
#   --timeout 300   --test-timeout 600
#   --hitl, --use-dockerfile, --custom-plan, --use-uv  all off
\end{lstlisting}

\subsection{SWE-agent}
\label{app:baseline-sweagent}

SWE-agent 1.1.0 under its own \texttt{config/default.yaml} harness. The
effective configuration is reproduced below with the runtime overrides folded
in. Its two task-carrying templates are ported from Repo2Run, with
\texttt{\{\{task\_details\}\}} rebound to the variable SWE-agent provides and
Repo2Run's trailing shell-prompt markers dropped, since this arm does not use
their system template.

\begin{lstlisting}[style=promptstyle]
# env: OPENROUTER_PROVIDER=DeepInfra    # pins the upstream provider behind OpenRouter
agent:
  model:
    name: openrouter/deepseek/deepseek-v4-flash-0731
    per_instance_cost_limit: 2.0
    per_instance_call_limit: 100
    temperature: 0.2                    # SWE-agent default is 0.0
    completion_kwargs:
      extra_body:
        thinking:
          type: disabled                # DeepSeek v4 enables reasoning by default
  templates:
    system_template: |-
      You are a helpful assistant that can interact with a computer to solve tasks.
    instance_template: |-
      We're currently setting up the environment for the following task. Here are the
      details:

      TASK:
      {{problem_statement}}

      INSTRUCTIONS:
      Now, you'll carry out this task on your own. Your terminal session has begun in
      the repository's root directory. Use the provided commands and any bash commands
      you need. Edit and check files as needed.

      The goal is to generate a Dockerfile that can successfully build and run the
      tests in the repository using the command "pytest --collect-only -q".

      NOTE:
      1. The repository is cloned into /repo.
      2. The Dockerfile should start with the following lines (if the base image is
         python:3.10 and repository is adamobeng/wddbfs):
      ```
      FROM python:3.10
      RUN pip install pytest
      RUN git clone https://github.com/adamobeng/wddbfs.git
      RUN mkdir /repo
      RUN git config --global --add safe.directory /repo
      RUN cp -r /wddbfs/. /repo && rm -rf /wddbfs/
      RUN rm -rf /wddbfs
      ```

      Your task includes:
      0. **Generate Dockerfile**: Create a file named "Dockerfile" in the root path
         (e.g, the absolute path is /Dockerfile).
      1. **Read Directory Structure**: Check the folder structure in the root directory.
      2. **Check the Configuration Files**: Inspect files like "requirements.txt",
         "setup.py", "setup.cfg", "Pipfile*", etc.
      3. **Determine Package Dependencies**: Handle dependencies and manage conflicting
         dependency versions.
      4. **Testing**: Ensure "pytest /repo --collect-only -q" runs without errors.
      5. **Generate Dockerfile**: Write necessary installation or setup steps determined
         from the inspection. If you finish run testing successfully, you should modify
         the Dockerfile in /Dockerfile.

      IMPORTANT TIPS:
      * Check the directory and files carefully.
      * Make sure Dockerfile commands are correct.
      * Use proper Dockerfile syntax and indentation.
      * Test the final Dockerfile by running "pytest /repo --collect-only -q".
    problem_statement_template: |-
      Set up the environment for the {{language}} repository {{full_name}}
      ({{repo_url}}) so that its test suite can be collected and run, and record the
      working setup as a Dockerfile at /Dockerfile.
    next_step_template: |-
      OBSERVATION:
      {{observation}}
    next_step_no_output_template: |-
      Your command ran successfully and did not produce any output.
  tools:
    execution_timeout: 600              # upstream default: 30
    total_execution_timeout: 7200       # upstream default: 1800
    env_variables:
      PAGER: cat
      MANPAGER: cat
      LESS: -R
      PIP_PROGRESS_BAR: 'off'
      TQDM_DISABLE: '1'
      GIT_PAGER: cat
    bundles:
      - path: tools/registry
      - path: tools/edit_anthropic
      - path: tools/submit              # default.yaml also ships review_on_submit_m; dropped
    registry_variables:
      USE_FILEMAP: 'true'
    enable_bash_tool: true
    parse_function:
      type: function_calling
env:
  deployment:
    type: docker
    image: python:3.10
    remove_container: false
    remove_images: false
  repo:
    type: local
    path: <repository at the pinned base commit>
\end{lstlisting}

\subsection{SWE-agent with DepGraph Context}
\label{app:baseline-sweagent-graph}

This arm is identical to SWE-agent (Appendix~\ref{app:baseline-sweagent}) except for one
context block appended to the end of its task prompt. The block inlines Graph2Env's initial
DepGraph and the build script compiled from it, both produced before Graph2Env's repair loop,
and the full graph is available in the container for the agent to query. The agent is told the
context may be incomplete, and it must still write its own Dockerfile.

\begin{lstlisting}[style=promptstyle,literate={—}{{\textemdash}}1]
ADDITIONAL CONTEXT. Produced before your session, for reference, in this container under
/g2e/ (these files are not available when your Dockerfile is built):
- /g2e/setup.sh              setup commands compiled for base image <base image>
- /g2e/runtime_handoff.json  environment the compiled commands expect at test time
- /g2e/depgraph.json, /g2e/depgraph_full.json (every package leaf), /g2e/depgraph_advisory.txt
  An advisory dependency graph of this repository, derived from static analysis: what it
  needs for environment setup. It is advisory, not ground truth — it may be incomplete or
  wrong, so verify before relying on it. Query the JSON with grep or python3 -c when the
  advisory below is not specific enough; these files can be large, so pull out the nodes you
  need rather than printing one whole.

<depgraph_advisory.txt: goal, unsatisfied requirements with evidence and candidate fixes,
 summary of satisfied requirements>

<setup.sh: the compiled build script>
\end{lstlisting}

\subsection{Claude Code}
\label{app:baseline-claudecode}

Claude Code is driven as a CLI inside a container, working as a non-root user
with \texttt{sudo}. The repository is placed at \texttt{/testbed}
and the deliverable is a Dockerfile written to \texttt{/testbed/Dockerfile.gen}.

\begin{lstlisting}[style=promptstyle]
# env: ANTHROPIC_BASE_URL=https://api.deepseek.com/anthropic
#      ANTHROPIC_API_KEY=<key>
#      DISABLE_AUTOUPDATER=1

claude -p \
  --model sonnet \
  --max-turns 100 \
  --permission-mode bypassPermissions \
  --output-format stream-json --verbose --include-partial-messages
\end{lstlisting}

\begin{lstlisting}[style=promptstyle]
We're currently setting up the environment for the following task. Here are the details:

TASK:
Set up the environment for the Python repository {full_name} (https://github.com/{full_name})
so that its test suite can be collected, and record the working setup as a Dockerfile at
/testbed/Dockerfile.gen.

INSTRUCTIONS:
Now, you'll carry out this task on your own. Your session has begun in the repository's root
directory. Use any bash commands you need. Edit and check files as needed.

The goal is to generate a Dockerfile that can successfully build and run the tests in the
repository using the command "pytest --collect-only -q".

NOTE:
1. The repository is cloned into /testbed.
2. The Dockerfile should start with the following lines (if the base image is python:3.11 and
   the repository is adamobeng/wddbfs):
```
FROM python:3.11
RUN pip install pytest
RUN git clone https://github.com/adamobeng/wddbfs.git /testbed
WORKDIR /testbed
```
3. The build runs as ROOT from an EMPTY image: drop every `sudo` prefix from the commands you
   write into the Dockerfile, and do not rely on any file from this container.
4. Install into the system Python -- do NOT create a virtualenv, since the grader runs the
   system python3.
Your task includes:
0. **Generate Dockerfile**: Create the Dockerfile at /testbed/Dockerfile.gen.
1. **Read Directory Structure**: Check the folder structure in the root directory.
2. **Check the Configuration Files**: Inspect files like "requirements.txt", "setup.py",
   "setup.cfg", "Pipfile*", etc.
3. **Determine Package Dependencies**: Handle dependencies and manage conflicting dependency
   versions.
4. **Testing**: Ensure "python -m pytest /testbed --co -q" runs without errors.
5. **Generate Dockerfile**: Write necessary installation or setup steps determined from the
   inspection. If you finish run testing successfully, you should modify the Dockerfile in
   /testbed/Dockerfile.gen.

IMPORTANT TIPS:
* Check the directory and files carefully.
* Make sure Dockerfile commands are correct.
* Use proper Dockerfile syntax and indentation.
* Test the final Dockerfile by running "python -m pytest /testbed --co -q".
\end{lstlisting}

\section{Metrics Protocol}
\label{app:metrics-protocol}

Every method is scored by the same harness, which measures the Dockerfile the method
produces rather than any result the method reports.

\paragraph{Rebuild.}
Each Dockerfile is rebuilt from scratch in a fresh container, with the repository
checked out at its pinned commit under \texttt{/testbed}. A repository whose image
fails to build scores 0 on both metrics and remains in the denominator.

\paragraph{EBSR gate.}
Test collection runs from \texttt{/testbed} with no path argument, so the
repository's own \texttt{testpaths} configuration applies. Following
Repo2Run~\citep{hu2025repo2run}, the gate passes on exit code 0 (tests collected).

\begin{lstlisting}[style=promptstyle]
python -m pytest --collect-only -q --disable-warnings
\end{lstlisting}

\paragraph{ESSR run.}
The full test suite is then executed in the same container, with a per-test timeout
of 120\,s and a 1800\,s limit on the whole run. Passing tests are counted from the
JUnit report, and $\mathrm{ESSR}_r=\min(\mathrm{passed}_r,D_r)/D_r$ is averaged over
repositories (Section~\ref{sec:metrics}).

\begin{lstlisting}[style=promptstyle]
python -m pytest -q --continue-on-collection-errors \
  --junit-xml=logs/junit.xml \
  --timeout=120 --timeout-method=signal
\end{lstlisting}

\section{Dataset Corpus}
\label{app:gold-counts}

{\small
\setlength{\tabcolsep}{0.6em}
\begin{longtable}{llr}
\caption{Benchmark repositories with their pinned commit and gold test count (Section~\ref{sec:benchmark}).}
\label{tab:unified-corpus}\\
\toprule
\rowcolor{gray!15}
Repository name & Commit sha & \# Tests \\
\midrule
\endfirsthead
\multicolumn{3}{l}{\emph{Table~\ref{tab:unified-corpus} -- continued from previous page}}\\[2pt]
\toprule
\rowcolor{gray!15}
Repository name & Commit sha & \# Tests \\
\midrule
\endhead
\midrule
\multicolumn{3}{r}{\footnotesize\emph{Continued on next page}}\\
\endfoot
\bottomrule
\endlastfoot
adamchainz/django-mysql & d6f050 & 739 \\
aiidateam/aiida-core & 9cff5f & 3,741 \\
alliander-opensource/weather-provider-api & 70f5d3 & 129 \\
anthropics/anthropic-sdk-python & d2f654 & 4,217 \\
anthropics/claude-quickstarts & 882638 & 280 \\
apiflask/apiflask & d71d64 & 376 \\
ArchipelagoMW/Archipelago & e6e0bc & 20,943 \\
AsyncFuncAI/deepwiki-open & d92819 & 109 \\
avaiga/taipy-core & 62f0d8 & 1,241 \\
Azure/azure-cli & 889dcc & 5,796 \\
Azure/azure-iot-ops-cli-extension & 6541bb & 7,021 \\
baserow/baserow & 0621bc & 11,730 \\
basxsoftwareassociation/basxconnect & a616dd & 18 \\
BeehiveInnovations/pal-mcp-server & 7afc7c & 886 \\
beeware/briefcase & 448b58 & 3,062 \\
benadida/helios-server & c7ed06 & 51 \\
benthayer/git-gud & d00a04 & 172 \\
BerriAI/litellm & a8979f & 65,652 \\
biothings/biothings.api & 657356 & 313 \\
brightway-lca/brightway2-io & e315f0 & 199 \\
bruin-data/ingestr & 1a980e & 19 \\
catalystneuro/roiextractors & 4d98f3 & 475 \\
censys/censys-python & d5533d & 450 \\
centre-for-humanities-computing/dacy & d990de & 16 \\
Checkmk/checkmk & a92883 & 24,391 \\
chopratejas/headroom & 35b115 & 13,158 \\
cityofzion/neo3-boa & 58d33f & 2,102 \\
Cloud-CV/EvalAI & 1c5960 & 1,822 \\
co-me-tokens/CoMe & 5c7cfb & 223 \\
coderamp-labs/gitingest & 4e259a & 160 \\
columnflow/columnflow & ad0477 & 72 \\
conan-io/conan-package-tools & decb63 & 127 \\
containers/podman-compose & 8156ab & 736 \\
convexengineering/gpkit & bea123 & 286 \\
copier-org/copier & 454ec4 & 1,148 \\
crystaldba/postgres-mcp & 07eb32 & 216 \\
crytic/slither & 5e78b8 & 7,296 \\
D4Vinci/Scrapling & 8e7bc9 & 768 \\
datactive/bigbang & 42568c & 73 \\
datamol-io/datamol & 031238 & 282 \\
devopness/devopness & ea00b9 & 84 \\
diefenbach/django-lfs & 3e6e96 & 484 \\
django-oauth/django-oauth-toolkit & 74b100 & 557 \\
Donkie/Spoolman & eafbc6 & 223 \\
dtmilano/androidviewclient & 8738d3 & 320 \\
eastsidepreparatoryschool/epschedule & 76cc6b & 13 \\
ecds/readux & 3eaf18 & 320 \\
ellmetha/django-machina & e34c8b & 462 \\
facebookresearch/hydra & 2e682d & 2,615 \\
fastapi/typer & 7fb978 & 1,379 \\
feast-dev/feast & f296d4 & 2,644 \\
fhempy/fhempy & 2c61a6 & 81 \\
folio-fse/folio\_migration\_tools & 4e994d & 450 \\
fonttools/fontbakery & a51772 & 918 \\
frappe/press & 73a2a4 & 846 \\
gip-inclusion/les-emplois & 816cac & 6,099 \\
google/nsscache & e0f270 & 306 \\
google/trax & 145147 & 3,492 \\
GoogleCloudPlatform/gcpdiag & 21d367 & 1,418 \\
googlecloudplatform/gsutil & 767984 & 1,575 \\
GoogleCloudPlatform/PerfKitBenchmarker & bd08f2 & 2,563 \\
GoogleCloudPlatform/ramble & 9ac589 & 2,885 \\
grycap/im & 096005 & 330 \\
guardrails-ai/guardrails & d85ea9 & 568 \\
gymrek-lab/TRTools & 99d7fb & 233 \\
hhursev/recipe-scrapers & 4788af & 184 \\
hitsz-ids/synthetic-data-generator & f36845 & 177 \\
huggingface/transformers & 031cfc & 167,927 \\
ibm/ibmi-bob & d74105 & 17 \\
idaholab/civet & 282f2c & 462 \\
idank/explainshell & e961c8 & 559 \\
IFRCGo/go-api & bfeec0 & 333 \\
insistence/RuoYi-Vue3-FastAPI & 726883 & 1,479 \\
intake/intake & 7cc9d8 & 688 \\
iss-mimic/mimic & d3d1b7 & 3 \\
jacebrowning/memegen & 40d586 & 254 \\
jackdewinter/pymarkdown & 446fbb & 6,930 \\
jaehyeon-kim/flink-demos & fe228c & 33 \\
jageo/lobsterpy & 55d8d2 & 135 \\
jaraco/keyring & afe0ab & 242 \\
jazzband/tablib & 212126 & 157 \\
jhao104/proxy\_pool & 9cc0ca & 248 \\
johnsnowlabs/nlu & 506860 & 252 \\
JonBunator/Enterr & 3fa9be & 38 \\
karlicoss/hpi & 35dd5d & 147 \\
karlicoss/promnesia & 5136f0 & 168 \\
kittinan/spotify-github-profile & 0f605f & 134 \\
lhotse-speech/lhotse & f9fb18 & 1,974 \\
librephotos/librephotos & 0a9e63 & 125 \\
lightkurve/lightkurve & a8f26d & 429 \\
m0wer/joinmarket-ng & a54d74 & 536 \\
mad-lab-fau/biopsykit & 2ad99f & 1,434 \\
marimo-team/marimo & 537b23 & 1,087 \\
markqvist/reticulum & 6ded42 & 68 \\
materialsvirtuallab/monty & 0c4abd & 127 \\
mckinsey/vizro & 192687 & 2,500 \\
mdolab/openaerostruct & 0448bf & 182 \\
MemTensor/MemOS & de8069 & 946 \\
menpo/menpo & bf08b9 & 819 \\
meta-pytorch/torchrec & b30d3a & 3,222 \\
microsoft/markitdown & e144e0 & 374 \\
microsoft/torchgeo & f2e178 & 2,578 \\
microsoftgraph/msgraph-sdk-python-core & e36c17 & 26 \\
Mirascope/mirascope & cd6bf0 & 326 \\
miurahr/aqtinstall & ac7da3 & 367 \\
mlflow/mlflow & 6dcba5 & 21,221 \\
mne-tools/mne-lsl & 03d598 & 168 \\
Modalities/modalities & 4aa2e8 & 118 \\
modelscope/AgentJet & 25118b & 63 \\
mollie/mollie-api-python & 957b44 & 264 \\
mozilla/addons-server & 6ff48b & 9,098 \\
mpmath/mpmath & ee1a81 & 2,318 \\
mvt-project/mvt & f5b0a3 & 165 \\
netneurolab/netneurotools & 49f83c & 205 \\
neurogym/neurogym & ee0a14 & 28 \\
NewFuture/DDNS & e2e055 & 912 \\
nginx-proxy/nginx-proxy & a64c03 & 469 \\
Nikita-Filonov/ai-review & e3e19d & 828 \\
NikolasMarkou/dl\_techniques & 31a385 & 34,754 \\
Nitrokey/pynitrokey & 0f3070 & 190 \\
nixtla/neuralforecast & 90b6fd & 1,226 \\
nomadkaraoke/karaoke-gen & ecd114 & 6,545 \\
NovaSky-AI/SkyRL & 53b115 & 2,749 \\
numba/llvmlite & 5c181f & 380 \\
nutti/fake-bpy-module & eea267 & 126 \\
NVIDIA-NeMo/RL & b03da0 & 2,601 \\
oarriaga/paz & cede0a & 591 \\
observatorycontrolsystem/observation-portal & 31c2dd & 883 \\
online-ml/river & 6995f1 & 3,432 \\
onyx-dot-app/onyx & f4b2f6 & 14,119 \\
open-cas/open-cas-linux & ba5bdf & 6,197 \\
OpenCTI-Platform/connectors & 0de8b8 & 9,939 \\
OpenDCAI/DataFlow & d70f9e & 161 \\
opengisch/pum & 5b458a & 195 \\
pajbot/pajbot & a0d969 & 89 \\
pastas/pastas & eeb8e2 & 173 \\
pathintegral-institute/mcpm.sh & 6a92e5 & 298 \\
peering-manager/peering-manager & 4dcdd6 & 882 \\
perplexityai/perplexity-py & 299016 & 806 \\
piccolo-orm/piccolo & 17c0a8 & 823 \\
polarsource/polar & 97e943 & 5,749 \\
PostHog/posthog & 682044 & 77,642 \\
pre-commit/pre-commit & 1553b4 & 820 \\
pretalx/pretalx & 64e726 & 1,391 \\
pretix/pretix & 943b31 & 5,995 \\
pycontribs/jira & bf7306 & 311 \\
pypa/distutils & 0de29d & 478 \\
pytest-dev/pytest-xdist & c7b4f6 & 257 \\
python-control/python-control & dbc998 & 3,625 \\
python-markdown/markdown & 33359f & 1,149 \\
python-poetry/poetry & 5bab98 & 1,695 \\
python-websockets/websockets & ff4869 & 2,248 \\
qiboteam/qibocal & 0625d9 & 1,134 \\
Qiskit/qiskit & 655dfb & 64,558 \\
Quantum-Accelerators/quacc & f6bc96 & 335 \\
resend/resend-python & fbc422 & 695 \\
revoxhere/duino-coin & 2d3d3d & 130 \\
rfsbraz/deleterr & b64a08 & 880 \\
rq/rq & eacec8 & 690 \\
rstcheck/rstcheck & b8ddb0 & 62 \\
sassoftware/python-sasctl & ebd0a6 & 610 \\
scitools/iris-grib & e6e3d5 & 581 \\
scylladb/scylla-cluster-tests & 4def67 & 7,451 \\
scylladb/sphinx-scylladb-theme & 4e3917 & 35 \\
sgl-project/sglang-jax & dc84c1 & 3,557 \\
sgl-project/SpecForge & ed64d2 & 1,257 \\
shotgun-sh/shotgun & 4d344d & 2,503 \\
simplistix/testfixtures & 608b05 & 1,467 \\
skrub-data/skrub & 0d69d9 & 1,571 \\
sooperset/mcp-atlassian & 562f92 & 2,741 \\
spacetelescope/pysiaf & 4395e6 & 32 \\
stan-dev/cmdstanpy & aa7555 & 286 \\
sublimelinter/sublimelinter & 621ba9 & 1,642 \\
supabase/supabase-py & 6f522d & 753 \\
TencentBlueKing/bk-lite & c7fd93 & 39,652 \\
TencentBlueKing/blueking-apigateway & 87bc68 & 4,646 \\
TencentBlueKing/blueking-dbm & ab940a & 3,581 \\
tensorflow/model-optimization & 38b201 & 1,725 \\
testcontainers/testcontainers-python & 1571ff & 731 \\
tgoai/tgo & 995da4 & 102 \\
The-OpenROAD-Project/OpenLane & e0d2e6 & 79 \\
theislab/cellrank & 607491 & 1,151 \\
thenewboston-developers/thenewboston-Backend & 9ff731 & 90 \\
timbrel/gitsavvy & da8f3c & 383 \\
tinygrad/tinygrad & 96b94b & 5,476 \\
trailofbits/algo & f6931d & 129 \\
ucoproject/uco & 9f1683 & 46 \\
unit8co/darts & 4d5a9a & 10,915 \\
uploadcare/pyuploadcare & 114d17 & 267 \\
uriyyo/fastapi-pagination & e00623 & 1,732 \\
valory-xyz/trader & 64fa99 & 18 \\
wecode-ai/Wegent & f296d3 & 5,050 \\
wilcoxjay/mypyvy & 3a055d & 8 \\
wireless-innovation-forum/spectrum-access-system & 928c31 & 369 \\
wpi-lnl/lnldb & 6384c0 & 317 \\
xuwei95/ezdata & 97243e & 291 \\
yihong0618/bilingual\_book\_maker & 3f7fc1 & 2,294 \\
yourlabs/django-autocomplete-light & 45a9ff & 84 \\
zostera/django-bootstrap3 & a6565b & 64 \\
zscole/gru & 67c9b6 & 773 \\
\end{longtable}
}

\end{document}